\documentclass[%
 reprint,
 amsmath,amssymb,
 aps,
 prd,
floatfix,
showkeys
]{revtex4-1}

\usepackage[USenglish]{babel}
\usepackage{slashed}
\usepackage{graphicx}
\usepackage{dcolumn}
\usepackage{bm}
\usepackage{amssymb}
\usepackage{amsmath}
\usepackage{graphicx}
\usepackage{epsfig}
\usepackage{rotating}

\usepackage[utf8]{inputenc} 
\usepackage{amsmath} 
\usepackage{array}
\usepackage{float}
\usepackage{color}
\usepackage[usenames,dvipsnames]{xcolor}
\usepackage{epstopdf}
\usepackage{natbib}
\usepackage{graphicx}
\usepackage{lineno}
\usepackage{slashed}
\usepackage{color}
\usepackage{hyperref}
\usepackage{float}
\usepackage{mathtools,slashed}
\usepackage{multirow}

\newcommand{\be}{\begin{equation}}
\newcommand{\ee}{\end{equation}}
\newcommand{\bea}{\begin{eqnarray}}
\newcommand{\eea}{\end{eqnarray}}
\newcommand{\beas}{\begin{eqnarray*}}
\newcommand{\eeas}{\end{eqnarray*}}

\newcommand{\bse}{\begin{subequations}}
\newcommand{\ese}{\end{subequations}}

\begin{document}

\title{\bf Total angular momentum of water molecule and magnetic field interaction
}
\author{C. H.~Zepeda~Fern\'andez$^{1,2}$}
\email{Corresponding author, email:hzepeda@fcfm.buap.mx}
\author{C. A. L\'opez T\'ellez$^{2}$}
\author{Y. Flores Orea$^2$}
\author{E.~Moreno~Barbosa$^2$}

\address{
  $^1$Cátedra CONACyT, 03940, CdMx Mexico\\
  $^2$Facultad de Ciencias F\'isico Matem\'aticas, Benem\'erita Universidad Aut\'onoma de Puebla, Av. San Claudio y 18 Sur, Ciudad Universitaria 72570, Puebla, Mexico\\
}

\begin{abstract}
  \begin{center}
    \bf \large Abstract
  \end{center}
  One of the must important non-invasive techniques in medicine is the Magnetic Resonance Imaging (MRI), it is used to obtain information of the structure of the human body parts using three dimensional images. The technique to obtain these images is based by the emission of radio waves produced by the protons of the hydrogen atoms in water molecules when placed in a constant magnetic field after they interact with a pulsed radio frequency (RF) current, the spin of the protons are in a spin excited state. When the RF field is turned off, the MRI sensors are able to detect the energy released (RF waves) as the protons realign their spins with the magnetic field.\\
  We used a three particles model for the water molecule: the two protons from the hydrogen atoms move around the doubly negatively charged oxygen (unstructured),  to describe the total angular momentum. The energy levels from the water molecule are studied in presence of an uniform external magnetic field, which interacts with the proton's spin and orbital angular momentum. The energy is shifted and the degeneration is lifted. To illustrate the results, we provide numerical results for a magnetic field strength commonly used in MRI devices.
\end{abstract}


\maketitle

\section{Introduction}\label{1}
One of the must important non-invasive techniques in medicine to obtain information of the structure and composition of the human body parts using three dimentional images is the Magnetic Resonance Imaging (MRI)~\cite{mr1,mr2,mr3}. The images obtained are high resolution compared to other techniques that are invasive, as the Computed Tomography (CT)~\cite{ct1,ct2,ct3}, which requires  using x-rays~\cite{xr1,xr2,xr3}.  The procedure to obtain high resolution images is constantly development~\cite{mri1,mri2,mri3,mri4}.\\
The MRI scanner is a complex device~\cite{mri}, which needs a strong and constant magnetic field between 0.5 and 7~T. This magnetic field interacts with the spin of the in the hydrogen atoms that make up the water molecules, which make up the human body by 80\%~\cite{water}. Some spins are oriented parallel and some others anti-parallel with respect to the magnetic field direction. Then, the radio frequency coils (RFC)~\cite{mri} produce a radio frequency field (RF), which reverse the orientation of the spins (parallel becomes anti-parallel and vice versa). When, the RF is turned off, the spins return to their original orientation. In this transition, the protons produce radio waves that are detected by the MRI device and finally the image is produced~\cite{mri}. The protons and neutrons (from other atoms that make up other molecules in the human body) do not contribute to the magnetic interaction, because they are mostly spin-paired. On the other hand, the spin of the electrons that are not in spin-paired, also interact with the magnetic field, however, they radiate in microwaves, which are not detected by the RFC (see section II of~\cite{RMF}).\\\\
We use a simple model for the water molecule, which consists of a triangle shape at whose vertices are the two protons and the oxygen atom (doubly charged negative and unstructured), this shape is due to the electrostatic interaction, holding the distance between protons constant (151.05~pm) and the distance between the protons and the oxygen, also constant (95.60~pm)~\cite{we1}. In a previous work~\cite{RMF}, we showed the energy levels for the protons of the hydrogen atoms in the  water molecule, where it was also shown the breakdown of degeneracy caused by the interaction with an external constant magnetic field, for the quantum numbers $m_1$ and $m_2$ (the magnetic quantum number for each proton), nevertheless, the degeneration remains in $l_1$ and $l_2$ (the orbital angular momentum quantum number for each proton). In this previous work, the interaction between the spin of the protons and the constant external magnetic field was not considered, only the interaction with the orbital angular momentum was considered.\\
In the present work, we describe the interaction between the total angular momentum (${\bf \hat J=\hat L+\hat  S}$) of the water molecule with the MRI magnetic field. The model can be thought as a hydrogen atom, then, we make this analysis analogous to the anomalous Zeeman effect. The work is organized as follows: In Section~\ref{2} we show the states of the total angular momentum. 
In Section~\ref{3} we introduce a constant magnetic field to obtain the fine structure for our water molecule model and we find the total energy considering the total angular momentum. Finally, in Section~\ref{4} we discuss our results and conclude.

\section{Total angular momentum of water molecule}\label{2}
As already described the protons of the hydrogen atom of the water molecule,  are the particles of our analysis, due to their interaction with the magnetic field in the MRI study. It is well known that their spin is 1/2, then, the total angular momentum follows the formalism of angular momentum addition: the sum of both spins and the orbital angular momentum $l_1$ and $l_2$. As we know, the addition of angular momentum, it is carried out two by two. We first add $l_i$ and their respective  spin 1/2, for both protons, $i=1, 2$. For this work, we use $i$ to refer to proton 1 or proton 2. The case $l_i=0$ implies that there is no angular momentum and the total angular momentum is purely spin angular momentum. For the case of $l_i>0$, the resulting angular momentum $j_i$, only  has two possibilities $j_i=l_i\pm1/2$. Then, the states are giving by:

\begin{equation}\label{jimi}
  \begin{split}
    \displaystyle |j_i;m_i>_{l_{i}}=&\Big|l_i\pm\frac{1}{2};m_i\Big>_{l_{i}}=\\
    &\sqrt{\frac{l_i\mp m_i+\frac{1}{2}}{2l_i+1}}\Big|l_i,\frac{1}{2};m_i+\frac{1}{2},-\frac{1}{2}\Big>\pm\\
    & \sqrt{\frac{l_i\pm m_i+\frac{1}{2}}{2l_i+1}}\Big|l_i,\frac{1}{2};m_i-\frac{1}{2},\frac{1}{2}\Big>,
  \end{split}
\end{equation}

where, $m_i=-j_i,-j_i+1...,j_i-1,j_i$ and $|l_i,\frac{1}{2};m_i\pm \frac{1}{2},\mp \frac{1}{2}>=|\l_i;m_i\pm \frac{1}{2}>|\frac{1}{2};\mp \frac{1}{2}>$ is the tensor product of the spatial part and the spin part. 
The spin part is

\begin{equation}
  \Big|\frac{1}{2};\frac{1}{2}\Big>=
  \begin{pmatrix}
    1 \\
    0 \\
  \end{pmatrix},
  \Big|\frac{1}{2};-\frac{1}{2}\Big>=
  \begin{pmatrix}
    0 \\
    1 \\
  \end{pmatrix}
\end{equation}

Then, we can write Eq~\ref{jimi} as follows:

\begin{equation}\label{j1j2}
  \begin{split}
    \Phi_{j_i=l_i\pm\frac{1}{2},m_i}(\theta_i,\phi_i)=&\frac{1}{\sqrt{2l_i+1}}\times\\
    & \begin{pmatrix}
        \pm\sqrt{l_i\pm m_i+\frac{1}{2}}Y_{l_i,m_i-\frac{1}{2}}(\theta_i,\phi_i)\\
        \sqrt{l_i\mp m_i-\frac{1}{2}}Y_{l_i,m_i+\frac{1}{2}}(\theta_i,\phi_i)
      \end{pmatrix}.
  \end{split}
\end{equation}

Where the $Y_{l_i,m_i\pm\frac{1}{2}}(\theta_i,\phi_i)$ are the spatial functions (spherical harmonics) for the water molecule~\cite{RMF}.\\
Finally, we add $j_1$ and $j_2$, where, it is obtained by four possible values of the total angular momentum  $j$, giving by the combinations:

\begin{enumerate}
\item $j_1=l_1-1/2$ and $j_2=l_2-1/2$
\item $j_1=l_1-1/2$ and $j_2=l_2+1/2$
\item $j_1=l_1+1/2$ and $j_2=l_2-1/2$
\item $j_1=l_1+1/2$ and $j_2=l_2+1/2$
\end{enumerate}

The total function is represented by  $|j;m>^{(l_1,l_2)}_{(j_{1},j_{2})}$ and  $m=-j,j+1...,j-1,j$. We can express the state $|j;m>^{(l_1,l_2)}_{(j_{1},j_{2})}$ in terms of the Clebsh-Gordan coefficients (${_{l_{1},l_{2}}}<j_1,j_2;m_1,m_2|j;m>$) as:

\begin{equation}
  \begin{split}
    \displaystyle |j;m>^{(l_1,l_2)}_{(j_{1},j_{2})}=&\sum_{m_1}\sum_{m_2} {_{l_{1},l_{2}}}<j_1,j_2;m_1,m_2|j;m>\times\\
    &|j_1,j_2;m_1,m_2>_{l_{1},l_{2}}.\\
  \end{split}
\end{equation}

Where
\begin{equation}\label{kets}
  |j_1,j_2;m_1,m_2>_{l_{1},l_{2}}=|j_1;m_1>_{l_{1}}\otimes|j_2;m_2>_{l_{2}}
\end{equation}

represents the Kronecker product. The wave function is giving by
\begin{equation}\label{totalfunction}
  \begin{split}
    \Psi_{l_1,l_2;m_1,m_2;\frac{1}{2},\pm\frac{1}{2}}(\theta_1,\phi_1,\theta_2,\phi_2)=&\\
    \Phi_{l_1\pm\frac{1}{2},m_1}(\theta_1,\phi_1)\Phi_{l_2\pm\frac{1}{2},m_2}(\theta_2,\phi_2).&
  \end{split}
\end{equation}

\subsection{Ground state of water molecule}\label{ground}
Fot the Clebsh-Gordan coefficients, it is well known that $<j_1,j_2;m_1,m_2|j;m>\neq0$ only when $m=m_1+m_2$. Other wise, when $m\neq m_1+m_2$ we have $<j_1,j_2;m_1,m_2|j;m>=0$, then, following this rules it can be obtained the states $|j;m>$ for a giving $j_1$ and $j_2$. Note, that the water molecule model is a fermion system, then, the protons follow the Pauli exclusion principle. For the ground state, we have $l_1=l_2=0$ and $m_1=m_2=0$, therefore, one of the spin protons must have $+1/2$ value and the other the  $-1/2$ value. The total angular momentum takes the value $j=0,1$ and the only two states allowed are

\begin{equation}\label{groundstate}
  \begin{split}
    \displaystyle |0;0>^{(0,0)}_{(\frac{1}{2},\frac{1}{2})}=\Big|\frac{1}{2},\frac{1}{2};-\frac{1}{2},\frac{1}{2}\Big>_{00} &\\
    \displaystyle |1;0>^{(0,0)}_{(\frac{1}{2},\frac{1}{2})}=\Big|\frac{1}{2},\frac{1}{2};\frac{1}{2},-\frac{1}{2}\Big>_{00}&
  \end{split}
\end{equation}


\subsection{First state of water molecule}\label{first}
The first excited state must occur with $l_i = 1$ and the other remains in $l_i = 0$. For this case all combinations of the quantum numbers are allowed. To exemplify, we take $l_1 = 1$, then,  $j_1=\frac{1}{2},\frac{3}{2}$;  which the six states can be obtained by Eq.~\ref{jimi}. The quantum numbers for the second proton  are $l_2=0$ and $s_2=\frac{1}{2}$, then $j_2=\frac{1}{2}$. We describe the two sub-spaces below.

\subsubsection{Space generated by $j_{1}=\frac{1}{2}$ and  $j_{2}=\frac{1}{2}$}
The total angular momentum j has two possible values: 1 and 0.
\begin{enumerate}
\item Case j=1: The three states are
  \begin{equation}
    \begin{split}
      |1,1\rangle^{(1,0)}_{(\frac{1}{2},\frac{1}{2})}=&\left|\frac{1}{2}, \frac{1}{2} ; \frac{1}{2}, \frac{1}{2}\right\rangle_{10} \\
      |1,0\rangle^{(1,0)}_{(\frac{1}{2},\frac{1}{2})}=&\frac{1}{\sqrt{2}}\left(\left|\frac{1}{2}, \frac{1}{2} ;-\frac{1}{2}, \frac{1}{2}\right\rangle_{10}+\left|\frac{1}{2}, \frac{1}{2} ; \frac{1}{2},-\frac{1}{2}\right\rangle_{10}\right) \\
      |1,-1\rangle^{(1,0)}_{(\frac{1}{2},\frac{1}{2})}=&\left|\frac{1}{2}, \frac{1}{2} ;-\frac{1}{2},-\frac{1}{2}\right\rangle_{10}
    \end{split}
  \end{equation}

\item Case j=0: There is only one state giving by
  \begin{equation}
    |0,0\rangle^{(1,0)}_{(\frac{1}{2},\frac{1}{2})}=\frac{1}{\sqrt{2}}\left(\left|\frac{1}{2}, \frac{1}{2}, \frac{1}{2},-\frac{1}{2}\right\rangle_{10}-\left|\frac{1}{2}, \frac{1}{2} ;-\frac{1}{2}, \frac{1}{2}\right\rangle_{10}\right)
  \end{equation}
\end{enumerate}
\subsubsection{Space generated by $j_{1}=\frac{3}{2}$ and  $j_{2}=\frac{1}{2}$}

The total angular momentum j has two possible values: 2 and 1.
\begin{enumerate}
\item Case j=2: The five states are
  \begin{equation}
    \begin{split}
      |2,2\rangle^{(1,0)}_{(\frac{3}{2},\frac{1}{2})}=&\left|\frac{3}{2}, \frac{1}{2} ; \frac{3}{2}, \frac{1}{2}\right\rangle_{10} \\
      |2,1\rangle^{(1,0)}_{(\frac{3}{2},\frac{1}{2})}=&\sqrt{\frac{3}{4}}\left|\frac{3}{2}, \frac{1}{2} ; \frac{1}{2}, \frac{1}{2}\right\rangle_{10}+\frac{1}{2}\left|\frac{3}{2}, \frac{1}{2} ; \frac{3}{2},-\frac{1}{2}\right\rangle_{10} \\
      |2,0\rangle^{(1,0)}_{(\frac{3}{2},\frac{1}{2})}=&\frac{1}{\sqrt{2}}\left(\left|\frac{3}{2}, \frac{1}{2} ;-\frac{1}{2}, \frac{1}{2}\right\rangle_{10}+\left|\frac{3}{2}, \frac{1}{2} ; \frac{1}{2},-\frac{1}{2}\right\rangle_{10}\right) \\
      |2,-1\rangle^{(1,0)}_{(\frac{3}{2},\frac{1}{2})}=&\frac{1}{2}\left|\frac{3}{2}, \frac{1}{2} ;-\frac{3}{2}, \frac{1}{2}\right\rangle_{10}+\sqrt{\frac{3}{4}}\left|\frac{3}{2}, \frac{1}{2} ;-\frac{1}{2},-\frac{1}{2}\right\rangle_{10} \\
      |2,-2\rangle^{(1,0)}_{(\frac{3}{2},\frac{1}{2})}=&\left|\frac{3}{2}, \frac{1}{2} ;-\frac{3}{2},-\frac{1}{2}\right\rangle_{10}
    \end{split}
  \end{equation}
\item Case j=1: There are three states giving by
  \begin{equation}
    \begin{split}
      |1,1\rangle^{(1,0)}_{(\frac{3}{2},\frac{1}{2})}=&\sqrt{\frac{3}{4}}\left|\frac{3}{2}, \frac{1}{2}, \frac{3}{2},-\frac{1}{2}\right\rangle_{10}-\frac{1}{2}\left|\frac{3}{2}, \frac{1}{2} ; \frac{1}{2}, \frac{1}{2}\right\rangle_{10} \\
      |1,0\rangle^{(1,0)}_{(\frac{3}{2},\frac{1}{2})}=&\frac{1}{\sqrt{2}}\left(\left|\frac{3}{2}, \frac{1}{2} ; \frac{1}{2},-\frac{1}{2}\right\rangle_{10}-\left|\frac{3}{2}, \frac{1}{2} ;-\frac{1}{2}, \frac{1}{2}\right\rangle_{10}\right) \\
      |1,-1\rangle^{(1,0)}_{(\frac{3}{2},\frac{1}{2})}=&\frac{1}{2}\left|\frac{3}{2}, \frac{1}{2} ;-\frac{1}{2},-\frac{1}{2}\right\rangle_{10}-\sqrt{\frac{3}{4}}\left|\frac{3}{2}, \frac{1}{2} ;-\frac{3}{2}, \frac{1}{2}\right\rangle_{10}
    \end{split}
  \end{equation}
\end{enumerate}

The energy of the water molecule is degenerated~\cite{RMF}. This degeneration is shifted when a magnetic field is considered. In Section~\ref{magneticfield}, we find the the shifted energy by the interaction of the total angular momentum with a constant magnetic field.

\subsection{State function for identical protons}
We mentioned that the water molecule is a fermion system, then, to consider the protons as identical particles, the wave function must be antisymmetric. The Eq.\ref{totalfunction} becomes

\begin{equation}
  \begin{split}
    \Psi_{l_1,l_2;m_1,m_2;\frac{1}{2},\pm\frac{1}{2}}(\theta_1,\phi_1,\theta_2,\phi_2)=&
    \\\frac{1}{\sqrt{2}}
    \begin{vmatrix}
      \Phi_{l_1\pm\frac{1}{2},m_1}(\theta_1,\phi_1) & \Phi_{l_2\pm\frac{1}{2},m_2}(\theta_1,\phi_1)\\
      \Phi_{l_1\pm\frac{1}{2},m_1}(\theta_2,\phi_2) & \Phi_{l_2\pm\frac{1}{2},m_2}(\theta_2,\phi_2) \\
    \end{vmatrix}
    &.
  \end{split}
\end{equation}

Where the $2\times2$ determinant is the Slater determinant.

\section{Interaction of water molecule with an external magnetic field}\label{3}
The principal assumption used in the MRI study is the interaction between the spin of the protons and the external magnetic field~\cite{PHMRI1,PHMRI2,PHMRI3}. The water molecule model used for this analysis  is like a hydrogen atom. Then, the presence of the water molecule in a constant magnetic field can be treated as the anomalous Zeeman effect, where the protons would be the analogous to the electron in the hydrogen atom. In this section we show the results of the spin-orbit coupling and the relativistic correction to obtain the fine structure. It will seen that these phenomenon are a perturbation, similar to the anomalous Zeeman effect. Finally, we show the total energy, considering the fine structure and the energy due to the interaction of the proton's spin and angular momentum with the magnetic field. 

\subsection{Spin-orbit coupling}
For the hydrogen atom, the spin-oribit coupling arises from the interaction between the electron's spin magnetic moment ($\displaystyle {\bf \hat\mu_{s_{e}}}=\frac{e}{m_ec}{\bf \hat S}$, where $e$ is the electric charge, $m_e$ is the electron mass and $c$ is the speed of light) and the magnetic field produced by the proton (${\bf B_p}$). This magnetic field is considering  from the electron's rest frame, the proton moves with a velocity $-v$. Then, the magnetic field is:

\begin{equation}\label{B}
  \displaystyle {\bf B_p}=\frac{1}{m_ec} {\bf E} \times  {\bf p}.
\end{equation}
Where, ${\bf p}$ is the electron's momentum and ${\bf E}$ is the Electric field produced by the proton. Finally, this electric field  can be expressed in terms of the Coulomb potential between the proton and the electron ($-e^2/r$) as
\begin{equation}\label{E}
  \displaystyle {\bf E}=\frac{1}{er} {\bf r}\frac{dV}{dr}.
\end{equation}
Where $r$ is the relative distance between the proton and the electron. Combining the equations~\ref{B} and \ref{E} it is possible to obtain the magnetic field in terms of the electron's angular momentum. Then, the interaction with the elctron's spin magnetic moment with the magnetic field is the known as the spin-orbit coupling  Hamiltonian

\begin{equation}\label{Hso}
  \displaystyle \hat H_{SO}^e=-\frac{1}{2m_e^2c^2}\frac{1}{r}\frac{d\hat V}{dr}{\bf \hat S \cdot \hat L}.
\end{equation}
The magnetic moment of the protons is $\displaystyle {\bf \hat \mu_{s_{p}}}=-g\frac{e}{2m_pc}{\bf \hat S}$, where $g\sim 5.585$~\cite{nist} is the gyromangentic ratio and $m_p$ is the proton  mass.  
Then, the Hamiltonian associated to the proton's spin-orbit is giving by

\begin{equation}
  \hat H_{SO}=-\frac{ge}{2m_p^2c}\frac{1}{r}\frac{dV}{dr}{\bf \hat S \cdot \hat L}.
\end{equation}

As we mentioned, for our water molecule model, the distance of the protons with the oxygen doubly charged negative, as well as the distance between the protons are constant, then $\displaystyle \frac{dV}{dr}=0$ and therefore $\hat H_{SO}=0$, i.e., there is no spin-orbit coupling contribution for our water molecule model.

\subsection{Relativistic correction}
The kinetic energy of one proton can be approximated as:
\begin{equation}\label{kin}
  \displaystyle  \hat T_i=\sqrt{\hat p_i^2c^2+m_p^2c^4}\approx \frac{\hat p_i^2}{2m_p}-\frac{\hat p_i^4}{8m_p^3c^2}+\cdots
\end{equation}
where $\hat p_i=-i\hbar\nabla_i$ is the momentum of the proton $i$. In spherical coordinates
\begin{equation}
  \displaystyle \nabla_i^2=\frac{1}{r_i}\frac{\partial^2}{\partial r_i^2}r-\frac{1}{\hbar^2r_i^2}\hat L_i^2.
\end{equation}
Where,
\begin{equation}
  \displaystyle \hat L_i^2=-\hbar^2\Big[\frac{1}{sin\theta_i}\frac{\partial}{\partial\theta_i}(sin\theta_i\frac{\partial}{\partial\theta_i})+\frac{1}{sin^2\theta_i}\frac{\partial^2}{\partial\phi_i^2}\Big]
\end{equation}
is the orbital angular momentum operator of the proton $i$. Once again, for our model, the $r_i$-coordinate is constant ($r_i=d_o$), then, the  momentum is purely orbital angular momentum: $\displaystyle \nabla^2=-\frac{1}{\hbar^2d_o^2}\hat L_i^2$.\\
The Hamiltonian is the sum of the kinetic and potential operators ($\hat H_i=\hat T_i+\hat V_i$), then, tacking until the second term from Eq.~\ref{kin}, we obtain

\begin{equation}\label{H0R}
  \displaystyle  \hat H_i=\frac{\hat p_i^2}{2m_p}-\frac{\hat p_i^4}{8m_p^3c^2}+\hat V_i
\end{equation}

where we define $\displaystyle \hat V_i=\frac{e^2}{2d_p}-\frac{e^2}{d_o}$. The total potential of the system is $\displaystyle \hat V=\hat V_1+\hat V_2 =\frac{e^2}{d_p}-\frac{2e^2}{d_o}$, for our water molecule model, whereby Eq.~\ref{H0R} can be written as:

\begin{equation}
  \displaystyle  \hat H_{i}=\hat H_{0_i}-\frac{\hat p_i^4}{8m_p^3c^2}=\hat H_{0_i}+\hat H_{R_i}.
\end{equation}

Where,
\begin{equation}\label{H2}
  \displaystyle  \hat H_{0_i}=\frac{\hat p_i^2}{2m_p}+\hat V_i
\end{equation}
is the Hamiltonian witout the magnetic field and the relativistic Hamiltonian is giving by
\begin{equation}
  \displaystyle \hat H_{R_i}=-\frac{\hat p_i^4}{8m_p^3c^2}.
\end{equation}

We can express $\hat p_i^4$ in terms of $\hat H_{0_i}$ and $\hat V_i$ using Eq.~\ref{H2}, obtaining

\begin{equation}\label{p4}
  \displaystyle \hat p_i^4=4m_p^2[\hat H_{0_i}^2-2\hat V_i\hat H_{0_i}+\hat V_i^2].
\end{equation}

We can obtain the energy values of $\hat H_{R_i}$ to calculate the expectation value:

\begin{equation}
  \begin{split}
    \displaystyle E_{R_i}=<\hat H_{R_i}>&=-\frac{1}{8m_p^3c^2}<\hat p_i^4>
  \end{split}
\end{equation}

The expectation value of $\hat p_i^4$ can be obtained using Eq.~\ref{p4}, then, we obtain  the relativistic energy:
\begin{equation}
  \displaystyle E_{R_i}=-\frac{1}{8m_p^3c^2}<\hat p_i^4>=-\frac{1}{2m_pc^2}[{E_{0_i}}^2-2V_iE_{0_i}+V_i^2].
\end{equation}
Where $E_{0_i}$ is the energy without magnetic field~\cite{RMF}:

\begin{equation}
  \displaystyle 
  E_{0_i}=\frac{\hbar^2}{2m_pd_0^2}l_i(l_i+1)+V_i.
\end{equation}

The total energy without magnetic field is $E_0=E_{0_1}+E_{0_2}$. Finally, the total relativistic energy is 
\begin{equation}\label{el1l2}
  \begin{split}
    \displaystyle E_{R}= &E_{R_1}+E_{R_2}\\
    &=-\frac{1}{2m_pc^2}[E_{0_1}^2+E_{0_2}^2\\
      &-2V_1E_{0_1}-2V_2E_{0_2}+V_1^2+V_2^2].
  \end{split}
\end{equation}

It is simple to note, that the degeneration still persists in $m_1$ and $m_2$. This energy is considered a perturbation because it is of the order of $10^{-14}$~eV.

\subsection{The fine structure of water molecule}
The fine structure is obtained by adding the spin-orbit and the relativistic correction. As we showed, there is no contribution of the spin-orbit correction, then, the fine structure is the relativistic correction giving by Eq.~\ref{el1l2}.

\subsection{Interaction of the proton's spin with a constant external magnetic field}\label{magneticfield}
To complete this analysis, we consider the spin of the protons from our molecule water model and its interaction with an external magnetic field. The interaction of the orbital angular momentum ($\displaystyle {\bf \hat \mu_L}=-\frac{e}{2m_pc}{\bf \hat L}$) and spin angular momentum (${\bf \hat \mu_{s_{p}}}$) with the magnetic field is giving by the Hamiltonian:
\begin{equation}
  \hat H_{LSB}=-\frac{e}{2m_pc}[{\bf \hat L}+g{\bf \hat S}]\cdot {\bf B}.
\end{equation}
Which, considering that the magnetic field is parallel to the z direction as ${\bf B}=B{\bf\hat z}$.  we obtain that the Hamiltonian for the water molecule is

\begin{equation}\label{HLSB}
  \displaystyle \hat H_{LSB}=-\frac{eB}{2m_pc}[\hat L_{1_z}+\hat L_{2_z}+g(\hat S_{1_z}+\hat S_{2_z})].
\end{equation}

Then, the total Hamiltonian is giving by

\begin{equation}\label{H}
  \hat H=\hat H_0+\hat H_R+\hat H_{LSB}.
\end{equation}

The total energy is obtained by the expectation value of Eq.~\ref{H}. $E_0$ and $E_R$ are already known. Due to the lower order of magnitude of $E_R$, we not consider its numerical value for Eq~\ref{H}. $E_{LSB}=<\hat H_{LSB}>$ is obtained from Eq.~\ref{HLSB}. Then, the energy $E_{LSB}$ is giving by,

\begin{equation}
  E_{LSB} = -\frac{eB \hbar}{2m_p c}\Big[m_{l_1} + m_{l_2} +m_{s_1}g + m_{s_2}g\Big]
\end{equation}

\begin{figure}[htbp]
  \begin{center}
    \includegraphics[width=0.5\textwidth]{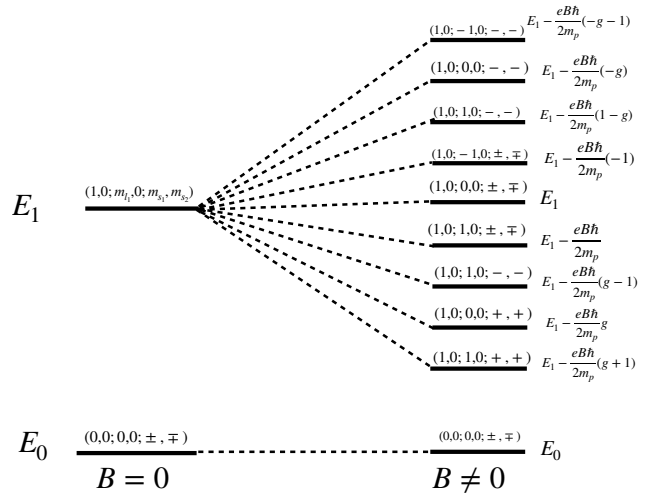}
  \end{center}
  \caption{Left: For the $B=0$ case. The first states are degenerated in the quantum numbers $m_{l_{2}},m_{s_{1}}$ and  $m_{s_{2}}$. Right: For the $B\neq0$ case, the degeneracy is lifted for all quantum numbers. The $\pm$ and $\mp$ indicates the interchange of spin.}
  \label{diagram}
\end{figure}

$m_{s_i}=\pm\frac{1}{2}$ and   $m_{l_i}=-l_i,-l_i+1,....,l_i-1,l_i$.\\
The water molecule model is a fermion system, then, the state of the protons  follow the Pauli exclusion principle. We use the notation ($l_1$, $l_2$; $m_1$, $m_2$; $m_{s_1}$, $m_{s_2}$) to represent the state of the system (Eq.~\ref{totalfunction}). $m_{s_{1}}$ and $m_{s_{2}}$ take the symbols $+$ and $-$, representing, $+\frac{1}{2}$ and $-\frac{1}{2}$, respectively. The ground state is giving by $(0,0;0,0;+,-)$, this state corresponds to $E_0$, as we showed in Subsection~\ref{ground}. Clearly, the degeneration persists with particle exchange. In Eq~\ref{groundstate}, the two states are orthogonal, however, they are actually representing an exchange of their spin and they are associated with the same energy.\\
The first excited state is giving by (1, 0; $m_1$, 0; $m_{s_{1}}$, $m_{s_{2}}$). We obtain nine different energy values for the lifted of $E_1$. In Figure~\ref{diagram} we show the lifted of $E_1$. In Table~\ref{values} are shown the respective values by considering B=7~T.  Then, the first exited state is for $(1,0;1,0;+,+)$, The second excited state occurs for $(1,0;1,0;+,+)$ and so on.

\begin{table}[htbp]
  \caption{Energy levels for the ground and first excited state of the protons in the water molecule in presence of a magnetic field of $7~T$.}
  \label{values}
  \centering
  \smallskip
  \begin{tabular}{| c | c | c |}
    \hline
    State & ($l_1,l_2;m_1,m_2,s_1,s_2$) & Energy (eV))   \\
    \hline
    \hline
    Ground & (0,0;0,0;$\pm$,$\mp$) & -23.4  \\
    \hline
    \hline
    \multirow{9}{1cm}{First}  & $( 1, 0; -1, 0; -, -)$ &  -23.3977255\\ \cline{2-3}
    & $(1,0;0,0;-,-)$ & -23.3977257  \\ \cline{2-3}
    & $(1,0;1,0;-,-)$ & -23.3977259  \\ \cline{2-3}
    & $(1,0;-1,0;\pm,\mp)$ & -23.3977267 \\ \cline{2-3}
    & $(1,0;0,0;\pm,\mp)$ & -23.3977270  \\ \cline{2-3}
    & $(1,0;1,0,\pm,\mp)$ & -23.3977272  \\ \cline{2-3}
    & $( 1,0;1,0;-,-)$ & -23.3977280  \\ \cline{2-3}
    & $(1,0;0,0;+,+)$ & -23.3977282 \\ \cline{2-3}
    & $(1,0;1,0;+,+)$ &  -23.3977284 \\ \cline{2-3}
    \hline

  \end{tabular}
\end{table}

\section{Discussion and conclusions}\label{4}
The working principle behind MRI devices is the interaction of the constant magnetic field with the spin of the protons (hydrogen nuclei) in the water molecule. In a previous work~\cite{RMF}, we calculated the function state of the protons, as well the energy, which is degenerated. This degeneracy is almost lifted when the water molecule is immersed in a constant magnetic field. However, the degeneracy persists with respect to the quantum numbers $l_1$ and $l_2$. By using the same water model, we complete this previous study to consider the spin of the protons. We obtained the total angular momentum of the water molecule, which correspond to the sum of the two orbital angular  momentum of each proton ($l_1$ and $l_2$) and their respective spin, i.e., the sum of four angular momenta. To consider the spin part, this wave function is described by the six quantum numbers $l_1$, $l_2$, $m_1$, $m_2$, $m_{s_1}$ and $m_{s_2}$, where we showed that the energy is degenerated.  This degeneracy is lifted when the water molecule is immersed in a constant magnetic field. Comparing the water molecule model we used to the hydrogen atom, it was possible to make an analogy with the anomalous Zeeman effect. Due to the constant distance between the protons, the spin-orbit coupling is zero, then,  the fine structure depends on the relativistic correction, however, it is to lower ($10^{-14}$~eV) and it is negligible to the energy $E_0$. To introduce a constant magnetic field, the degeneracy is lifted. 
There are nine values of $E_1$ lifting, as we showed in Figure~\ref{diagram}. The energy values are not longer degenerated, because it depends of the spin configuration. The values differ in the range of $\mu$eV.\\
With this analysis, we showed the total angular momentum interaction of the water molecule  with a magnetic field. Obtaining the wave function as well the energy lifted,  hoping, these results can be used to improve the techniques to obtain the image from a MRI, having more finesse.

\begin{minipage}{.5\textwidth}
  \centering
  \bf \large References
\end{minipage}

\end{document}